\documentclass[%
 reprint,
 superscriptaddress,
 showpacs,preprintnumbers,
 amsmath,amssymb,
 aps,
 prl,
 longbibliography,
]{revtex4-1}

\usepackage{graphicx}
\usepackage{dcolumn}
\usepackage{bm}
\usepackage{color}
\usepackage{ulem}

\begin{document}

\preprint{APS/123-QED}

\title{
Partial disorder in an Ising-spin Kondo lattice model on a triangular lattice
}

\author{Hiroaki Ishizuka}
\affiliation{
Department of Applied Physics, University of Tokyo, Hongo, 7-3-1, Bunkyo, Tokyo 113-8656, Japan
}

\author{Yukitoshi Motome}%
\affiliation{
Department of Applied Physics, University of Tokyo, Hongo, 7-3-1, Bunkyo, Tokyo 113-8656, Japan
}

\date{\today}

\begin{abstract}
Phase diagram of an Ising-spin Kondo lattice model on a triangular lattice near 1/3-filling is investigated by Monte Carlo simulation. 
We identify a partially disordered phase with coexistence of magnetic order and paramagnetic moments, 
which was unstable in two-dimensional Ising models with localized spins only. 
The partial disorder emerges in the competing regime between a two-sublattice stripe phase and three-sublattice ferrimagnetic phase, at finite temperatures above an electronic phase separation. 
The peculiar magnetic structure accompanies a charge order and develops a gap in the electronic structure. 
The results manifest a crucial role of the nonperturbative interplay between spin and charge degrees of freedom in stabilizing the partial disorder. 
\end{abstract}

\maketitle

The antiferromagnetic (AF) Ising model on a triangular lattice is one of the most fundamental models for
geometrically frustrated systems. 
When the interaction is restricted to the nearest-neighbor (n.n.) pairs, frustration in each triangle
prevents the system from long-range ordering (LRO) down to zero temperature ($T$), and the ground state has
extensive degeneracy and associated residual entropy~\cite{Wannier1950,Houtappel1950,Husimi1950}.
The degenerate ground state is extremely sensitive to perturbations; e.g., further-neighbor interactions
can lift the degeneracy and induce a variety of LRO.

\begin{figure}
   \includegraphics[width=3.20in]{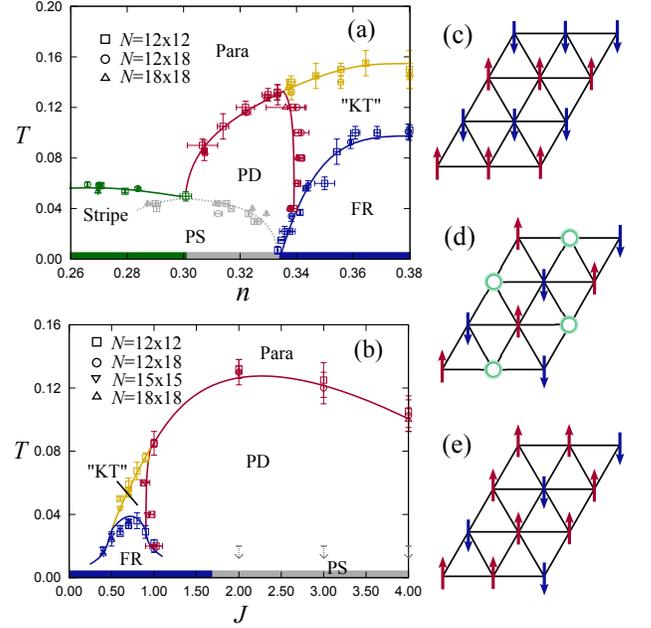}
   \caption{
   (color online). Phase diagram of the model~(\ref{eq:H}) (a) at $J=2$ while varying $n$ and (b) at $n=1/3$ while varying $J$.
   The symbols shows phase boundaries for the four phases: stripe phase, partially-disordered phase (PD),
   KT-like phase (KT), and ferrimagnetic phase (FR).  PS represents a phase separation.
   The lines are guides for the eyes.
   The colored strips at $T=0$ show the ground states obtained by comparing the energy of stripe and FR orders.
   The boundary between PD and PS in (b) is lower than $T=0.02$, and difficult to determine by MC calculations.
   The schematic pictures of the magnetic orders are given for (c) stripe, (d) PD, and (e) FR phases.
   The circles in (d) show paramagnetic sites.
   }
   \label{fig:phasediagram}
\end{figure}

An intriguing state emergent from the degenerate ground-state manifold is a partially disordered (PD) state. 
The PD state is peculiar coexistence of magnetically ordered moments and thermally-fluctuating paramagnetic moments.
Such possibility was first discussed in the presence of next n.n. ferromagnetic (FM) interaction~\cite{Mekata1977}. 
A mean-field study reported the existence of a three-sublattice PD phase with an AF ordering on the honeycomb
subnetwork and paramagnetic moments at the remaining sites [Fig.~\ref{fig:phasediagram}(d)]. 
Although such PD state was indeed observed in several Co compounds~\cite{Kohmoto1998,Niitaka2001}, subsequent 
Monte Carlo (MC) simulations indicated that, in purely two-dimensional systems, the PD appears at most as a quasi-LRO and that the transition from the high-$T$ paramagnetic phase is of Kosterlitz-Thouless (KT) type~\cite{Wada1982,Fujiki1983,Landau1983,Takayama1983,Fujiki1984,Takagi1995}.
The results suggest that, for establishing a PD LRO, it is indispensable to incorporate additional elements, such as a three-dimensional interlayer coupling~\cite{Todoroki2004}.

In this Letter, we explore the possibility of PD LRO in two dimensions when taking account of the coupling to itinerant electrons. 
Our study is partly motivated by the recent discovery of PD in metallic compounds with quasi-two-dimensional structure~\cite{Matsuda2012}.
The interplay between localized moments and itinerant electrons plays a crucial role in the following points. 
First of all, the kinetic motion of electrons induces effective interactions
known as the Ruderman-Kittel-Kasuya-Yosida (RKKY) mechanism~\cite{Ruderman1954,Kasuya1956,Yosida1957}. 
The long-ranged and oscillating nature of interaction drive keen competition between different magnetic states.
Furthermore, the change of magnetic states affects the electronic state in a self-consistent manner through the spin-charge coupling; the system can gain the energy by forming some particular electronic state associated with magnetic ordering.  
We note that similar PD states were recently discussed as a consequence of quantum spin fluctuation or hybirdization between itinerant and localized electrons~\cite{Motome2010,Hayami2011}.
In the present study, however, we will show that a minimal model, in which neither the quantum spin fluctuation nor hybridization operates, does exhibits a PD LRO.
Our results will provide deeper understanding of PD phenomena robustly observed in a broad range of materials, not only in the insulating localized spin systems but also the systems including itinerant electrons.

We consider a single-band Kondo lattice model on a triangular lattice with localized Ising spin moments.
The Hamiltonian is given by
\begin{eqnarray}
H = -t \! \sum_{\langle i,j \rangle, \sigma} \! ( c^\dagger_{i\sigma} c_{j\sigma} + \text{H.c.} ) + J \sum_{i}\sigma_i^z S_i.
\label{eq:H}
\end{eqnarray}
The first term represents hopping of itinerant electrons, where $c_{i\sigma}$ ($c^\dagger_{i\sigma}$) is the annihilation (creation) operator of an itinerant electron with spin $\sigma= \uparrow, \downarrow$ at $i$th site, and $t$ is the transfer
integral.
The sum $\langle i,j \rangle$ is taken over n.n. sites on the triangular lattice.
The second term is the onsite interaction between localized spins and itinerant electrons, where $\sigma_i^z$ represents the $z$-component of itinerant electron spin and $S_i = \pm 1$ denotes the localized Ising spin at $i$th site; $J$ is the coupling constant (the sign of $J$ does not matter in the present model). 
Hereafter, we take $t=1$ as the unit of energy, the lattice constant $a = 1$, and the Boltzmann constant $k_{\rm B} = 1$.
To investigate thermodynamic properties of this model, we adopted a MC simulation which is widely used for similar models~\cite{Yunoki1998}.
The calculations were conducted up to the system size $N=18^2$ under the periodic boundary conditions.
The thermal averages were calculated for typically 4300-9800 MC steps after 1700-5000 MC steps for thermalization. 

To describe the PD and ferrimagnetic (FR) states, here we use the pseudo-spin defined for each three-site unit cell,
\begin{eqnarray}
\tilde{\bf S}_m = 
\left(
\begin{array}{ccc}
\frac2{\sqrt6} & -\frac1{\sqrt6} & -\frac1{\sqrt6} \\
0              &  \frac1{\sqrt2} & -\frac1{\sqrt2} \\
\frac1{\sqrt3} &  \frac1{\sqrt3} &  \frac1{\sqrt3} \\
\end{array}
\right)
\left(
\begin{array}{c}
S_i  \\
S_j  \\
S_k  \\
\end{array}
\right),
\end{eqnarray}
and its summation ${\bf M} = \sum_m 3 \tilde{\bf S}_m/N$, where $m$ is the index for the three-site unit cells, and $(i,j,k)$ denote the three sites in the $m$th unit cell belonging to the sublattices (A,B,C), respectively~\cite{Fujiki1984}.
Then, the three-sublattice PD state [Fig.~\ref{fig:phasediagram}(d)] is characterized by a finite ${\bf M} = (M_x,M_y,M_z)$ parallel to $(\sqrt{3/2},1/\sqrt2,0)$, $(0,\sqrt2,0)$, or their threefold symmetric directions around the $z$-axis.
On the other hand, the three-sublattice FR state [Fig.~\ref{fig:phasediagram}(e)] is characterized by a finite ${\bf M}$ along $(\sqrt{2/3},\sqrt2,1/\sqrt3)$, $(2\sqrt{2/3},0,-1/\sqrt3)$, or their threefold symmetric directions around the $z$-axis.
Hence, the two states are distinguished by the azimuth of ${\bf M}$ in the $xy$-plane (as well as $M_z$). 
To parametrize the situation, we introduce $\psi = \tilde{M}_{xy}^3 \cos{6 \phi_M}$,
where $\phi_M$ is the azimuth of $\bf M$ in the $xy$ plane and $\tilde{M}_{xy} = 3M_{xy}^2/8$ ($M_{xy}^2 = M_x^2 + M_y^2$).
The parameter $\psi$ has a negative value and $\psi \to -\frac{27}{64}$ for the PD ordering, while it 
becomes positive and $\psi \to 1$ for the FR ordering; $\psi=0$ for both paramagnetic and KT phases in the bulk limit.
We calculate $\psi$, ${\bf M}$ and corresponding susceptibility, and spin entropy ${\cal S}$ (defined later) 
to map out the phase diagram.

Figure~\ref{fig:phasediagram}(a) shows the phase diagram around the electron density
$n=\sum_{i\sigma} \langle c_{i\sigma}^\dagger c_{i\sigma} \rangle/N=1/3$ at $J=2$ obtained by MC calculations. 
There are four dominant ordered phases ---the stripe, PD, FR, and KT phases, in addition to an electronic phase separation (PS).
In the relatively low density region $n \lesssim 0.30$, the stripe order with period two [Fig.~\ref{fig:phasediagram}(c)]
develops in the low $T$ region.
This is an interesting state, in which the magnetic order breaks the sixfold rotational lattice symmetry; Reflecting the lower symmetry, the electronic transport becomes spatially anisotropic.
On the other hand, in the higher density region $n \gtrsim 1/3$, the system exhibits the three-sublattice FR order [Fig.~\ref{fig:phasediagram}(e)] at low $T$. 
In this density region, we find another phase transition at a higher $T$ than the onset of FR order. 
As described below, we identify the intermediate phase as a KT-like state similar to the KT state discussed in the Ising models~\cite{Wada1982,Fujiki1983,Landau1983,Takayama1983,Fujiki1984}.
In the intermediate region $0.30 \lesssim n \lesssim 0.34$, the three-sublattice PD phase [Fig.~\ref{fig:phasediagram}(d)] emerges at finite $T$, whereas it is taken over by PS (or FR) at low $T$.
In the finite-$T$ region, the numerical data indicate that the PD state retains LRO, in sharp contrast to the KT region, as discussed below.

\begin{figure}
   \includegraphics[width=3.20in]{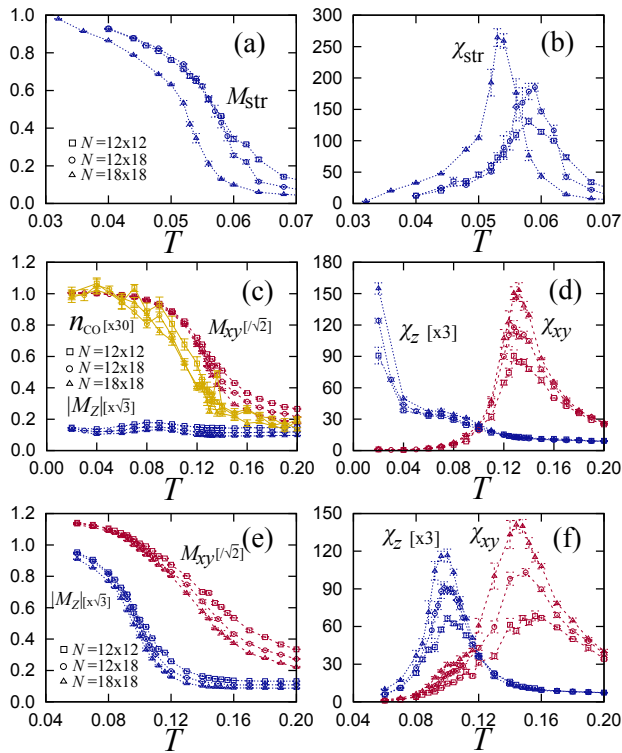}
   \caption{
   (color online).
   MC results for (a) $M_{\rm str}$, (b) $\chi_{\rm str}$, (c),(e) $M_{xy}$, $|M_z|$, and $n_{\rm CO}$,
   and (d),(f) $\chi_{xy}$ and $\chi_z$ for the system sizes $N=12\times12, 12\times18, 18\times18$.
   The data are at (a),(b) $n=0.27$, (c),(d) $n=1/3$, and (e),(f) $n=0.38$.
   }
   \label{fig:orderparam}
\end{figure}

Figure~\ref{fig:orderparam} shows typical $T$ dependences of physical quantities in each density region.
Figures~\ref{fig:orderparam}(a) and \ref{fig:orderparam}(b) present the results in the stripe region at $n=0.27$.
Figure~\ref{fig:orderparam}(a) shows the order parameter for stripe order, $M_{{\rm str}} = [ \sum_{{\bf q}^{*}_{\rm str}} \{S({\bf q}^{*}_{\rm str})/N\}^2 ]^{1/2}$, where $S({\bf q})$ is the spin structure factor of Ising spins and ${\bf q}^{*}_{\rm str}= (\pi,0), (\pm\frac12\pi,\frac{\sqrt3}2\pi)$ are the characteristic wave vectors of the stripe order;
Figure~\ref{fig:orderparam}(b) represents the corresponding susceptibility $\chi_{\rm str}$.
A phase transition to the stripe phase is signaled by a rapid increase of $M_{{\rm str}}$ and corresponding peak of
$\chi_{\rm str}$; the peak $T$ of $\chi_{\rm str}$, $T_c^{\rm (str)}$, for each system size is plotted in the phase diagram in Fig.~\ref{fig:phasediagram}(a). 

\begin{figure}
   \includegraphics[width=3.2in]{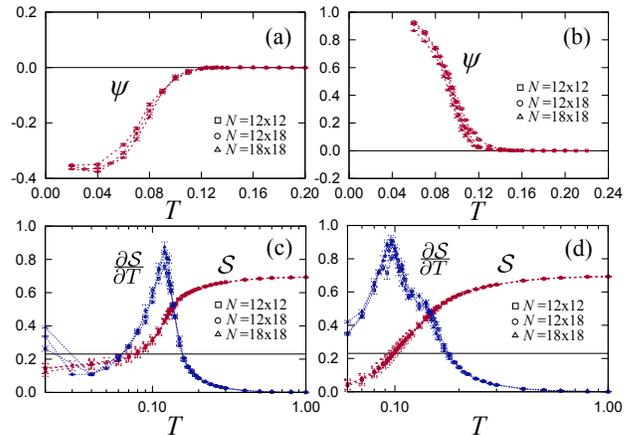}
   \caption{
   (color online).
   MC results for (a),(b) the $\psi$ parameter and (c),(d) spin entropy ${\cal S}$ and its derivative $\partial {\cal S}/\partial T$ for the system sizes $N=12\times 12, 12\times 18, 18\times18$.
   The data are at (a),(c) $n=1/3$ and (b),(d) $n=0.38$.
   $\cal S$ are calculated from numerical integration of $\partial {\cal S}/\partial T$, assuming ${\cal S}(T=1)=\log2$.
   }
   \label{fig:psi_seff}
\end{figure}

The results in the PD region for the pseudo-spin moments and the corresponding susceptibilities at $n=1/3$ 
are shown in Figs.~\ref{fig:orderparam}(c) and \ref{fig:orderparam}(d), respectively.
$M_{xy}$ shows a rapid increase around $T_c^{\rm (PD)} = 0.130(4)$ and approaches $\sqrt{2}$ at low $T$, whereas $|M_z|$ does not grow.
Correspondingly, $\chi_{xy}$ shows a divergent peak at $T_c^{\rm (PD)}$.
The low temperature phase is identified as the PD phase by the azimuth parameter $\psi$ shown in Fig.~\ref{fig:psi_seff}(a);
$\psi$ becomes negative below $T_{c}^{{\rm (PD)}}$ and approaches $-\frac{27}{64}$, as expected for the PD instability. 
At low $T$, the PD phase becomes unstable and is taken over by PS, which is determined by the jump of $n$ as a function of the chemical potential.

Figures~\ref{fig:orderparam}(e) and \ref{fig:orderparam}(f) show the results in the FR region at $n=0.38$.
The data indicate two successive transitions indicated by the peaks in $\chi_{xy}$ and $\chi_{z}$ at different $T$.
The peak of $\chi_z$ with corresponding increase of $|M_z|$ signals the phase transition to the FR phase at $T_c^{\rm (FR)} = 0.098(4)$.
On the other hand, at a higher $T_{\rm KT} = 0.146(4)$, only $M_{xy}$ changes rapidly, and correspondingly, $\chi_{xy}$ shows a peak.
$M_{xy}$, however, shows a noticeable system-size dependence even below $T_{\rm KT}$, in contrast with the results below $T_c^{\rm (PD)}$ in Fig.~\ref{fig:orderparam}(c).
Similar behavior was observed in the KT ordering in Ising spin systems~\cite{Takayama1983,Fujiki1984}.
Furthermore, $\psi$ does not show any clear anomaly at $T_{\rm KT}$, while it shows a sharp rise around $T_{c}^{{\rm (FR)}}$, as shown in Fig.~\ref{fig:psi_seff}(b).
The value of $\psi$ extrapolated to large $N$ converges to zero in the intermediate $T$ range.
This indicates that there is no sixfold symmetry breaking in $M_{xy}$ at $T_{\rm KT}$, as seen in the KT phase in the Ising spin models~\cite{Takayama1983}.
The extrapolated $\psi$ becomes finite below $T_c^{\rm (FR)}$ and approaches 1 as expected for the FR ordering.
Hence, we consider that the higher-$T$ transition is of KT type; the system exhibits two successive transitions
from the paramagnetic phase to the KT phase, and the KT phase to the low-$T$ FR phase. 

The PD phase 
in the intermediate $n$ exhibits characteristic features reflecting the presence of the paramagnetic moments.
Figures~\ref{fig:psi_seff}(c) and \ref{fig:psi_seff}(d) show the results of effective spin entropy ${\cal S}$~\cite{Udagawa2012} at $n=1/3$ and $0.38$, respectively.
The data for $\partial {\cal S}/\partial T$ are also shown, which give the specific heat associated with the Ising spins divided by $T$.
For $n=1/3$ [Fig.~\ref{fig:psi_seff}(c)], ${\cal S}$ shows a steep decrease around $T\sim T_c^{(\rm PD)}$, corresponding to the phase transition to the PD state.
However, it does not decrease to zero below $T_{c}^{{\rm (PD)}}$ and stays at a value close to $\frac13\log2$, the residual entropy expected for $1/3$ disordered moments in the ideal three-sublattice PD state~\cite{note_psi_S}.
In contrast, ${\cal S}$ in the FR region [Fig.~\ref{fig:psi_seff}(d)] shows little change at $T_{\rm KT}$, whereas a sharp decrease is seen at $T_c^{\rm (FR)}$;
${\cal S}$ becomes smaller than $\frac13\log2$ and approaches zero while decreasing $T$, as expected for the FR LRO.
The contrasting behavior depending on the presence and absence of the paramagnetic moments is also seen in the susceptibility $\chi_z$ shown in Figs.~\ref{fig:orderparam}(d) and \ref{fig:orderparam}(f).

We also note that the PD state accompanies concomitant charge modulation.
In Fig.~\ref{fig:orderparam}(c), we show the result for the charge order parameter $n_{\rm CO} = \{N({\bf q}^*_{\rm CO})/N\}^{1/2}$ at ${\bf q}^*_{\rm CO}=(-2\pi/3,2\pi/\sqrt3)$ [$N({\bf q})$ is the charge structure factor for the itinerant electrons].
The result shows a rapid increase of $n_{\rm CO}$ below $T_c^{\rm (PD)}$,
indicating that a three-sublattice charge order develops with PD 
(local charge density becomes poor at the paramagnetic sites).

Performing the analyses above in a wide range of parameters, we conclude that the system exhibits the three-sublattice PD LRO in the region near $n=1/3$ at finite $T$, while it turns into a KT-like quasi-LRO for larger $n$, as summarized in the phase diagram in Fig.~\ref{fig:phasediagram}(a). 
We also extend the analyses while changing $J$ and confirm that the PD state near $n=1/3$ remains stable in the wide range of $J$.
For instance, Fig.~\ref{fig:phasediagram}(b) shows the phase diagram at $n=1/3$ while varying $J$; 
The PD state appears for $J \gtrsim 1.0$, while the FR as well as KT state is stabilized for smaller $J$.

Let us discuss our results in comparison with the previous ones for the Ising spin models. 
As shown in Fig.~\ref{fig:phasediagram}(a), the PD state appears in the phase competing region between the stripe and FR phases. 
Although similar phase competition occurs in the Ising spin models while changing the second-neighbor interaction in the presence of the dominant AF interaction for n.n. pairs, PD does not appear in the form of LRO; 
the system shows, at most, a quasi-LRO with a KT transition~\cite{Wada1982,Fujiki1983,Landau1983,Takayama1983,Fujiki1984}. 
The PD LRO in our model, however, emerges in the distinct parameter regime from the KT region; 
it dominantly appears above the PS region in the phase competing regime. 
PS is an electronic origin associated with a finite jump of $n$ at the magnetic transition, and does not appear in the Ising spin models. 
Hence, our results indicate that the electronic degree of freedom opens a new window where the PD can be stabilized~\cite{note_PS}.

In addition, Fig.~\ref{fig:phasediagram}(b) shows that the PD phase appears only above a finite $J$ and is not stable down to the small $J$ limit.
This indicates that the RKKY interactions derived by the second-order perturbation in terms of $J$~\cite{Ruderman1954,Kasuya1956,Yosida1957} are not sufficient to explain the appearance of PD state. 
The result implies that the nonperturbative effect beyond the RKKY interactions takes a crucial role in stabilizing the PD LRO~\cite{note_n_ele}. 

\begin{figure}
   \includegraphics[width=3.20in]{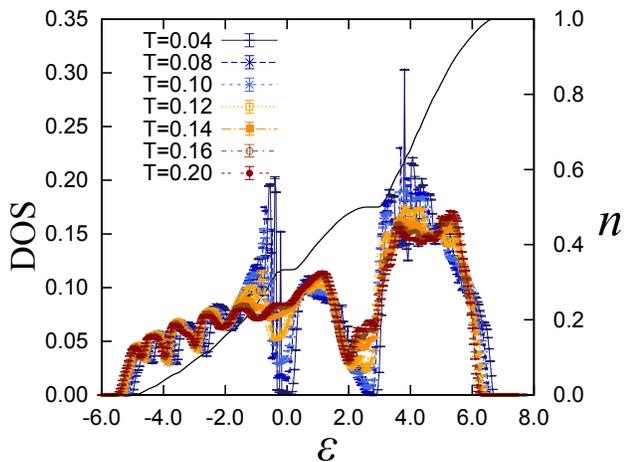}
   \caption{
   (color online).
   MC result for DOS of itinerant electrons at $n=1/3$ for $N=18\times 18$.
   The Fermi level is set at $\varepsilon = 0$.
   The solid curve shows the integrated DOS at $T=0.04$.
   }
   \label{fig:dos}
\end{figure}

Related to this point, we find that the PD state is stabilized by the change of the electronic structure.
Figure~\ref{fig:dos} shows the density of states (DOS) at $n=1/3$ while varying $T$.
The result shows the development of an energy gap at the Fermi level while decreasing $T$. 
The gap structure is always observed in the PD region, whereas such feature is not seen in the KT region. 
We note that a `mean-field' PD state with assuming the three-sublattice order of ($+m, -m, 0$) can open a charge gap for a sufficient $J$.
These results clearly indicate that the PD LRO is stabilized by an energy gain from the reconstruction of electronic state with gap opening.

To summarize, we have provided the convincing numerical results on the partial disorder in the Isin-spin Kondo lattice model on a triangular lattice.
Interestingly, our partially-disordered state emerges at finite temperatures dominantly above an electronic phase separation inherent to the itinerant model.
Furthermore, we found that the partial disorder develops with a charge gap in the electronic structure as well as charge ordering.
These results indicate that the nonperturbative interplay between spin and charge degrees of freedom is crucial for stabilizing the partial disorder.
Experimentally, recent studies on quasi-two-dimensional metallic oxides reported an indication of partial disorder. 
Our results will promote understanding of the interesting physics related to the peculiar coexistence of magnetic order
and paramagnetic moments in itinerant electron systems.

The authors are grateful to H. Kawamura, M. Matsuda, and H. Yoshida for fruitful discussions.
The authors also thank S. Hayami and T. Misawa for helpful comments.
Part of the calculations were performed on the Supercomputer Center, Insitute for Solid State Physics, University of Tokyo.
H.I. is supported by Grant-in-Aid for JSPS Fellows.
This research was supported by KAKENHI (No.19052008, 21340090, and 22540372), Global COE Program
``the Physical Sciences Frontier", the Strategic Programs for Innovative Research (SPIRE), MEXT, and the
Computational Materials Science Initiative (CMSI), Japan.


\begin{thebibliography}{99}
\bibitem{note_psi_S}       The small deviations of $\psi$ and ${\cal S}$ from their expected values are presumably ascribed to a spatial correlation between the disordered spins.
\bibitem{note_PS}          PS might be replaced by an inhomogeneous state by forming domains when taking account of additional elements such as impurities and Coulomb repulsions. Nevertheless the PD is expected to survive, because magnetically, it is an Ising-type order robust against (nonmagnetic) impurities, and electronically, it accompanies a charge order, which can be further stabilized by the long-range Coulomb repulsion.
\bibitem{note_n_ele}       We also conducted calculations by changing $n$, and found that the PD state always appears from a finite $J$.
\end{thebibliography}
\end{document}